# Two-state, Reversible, Universal Cellular Automata In Three Dimensions

Daniel B. Miller and Edward Fredkin

Carnegie Mellon University

West Coast Campus

Bldg 23, Nasa Research Park

Moffett Field, CA  94035

**Abstract**

A novel two-state, Reversible Cellular Automata (RCA) is described.  This three-dimensional RCA is shown to be capable of universal computation.  Additionally, evidence is offered that this RCA Is capable of universal construction.

**Acknowledgements**

The authors wish to thank Suresh Kumar Devanathan, who has provided valuable input on the issues discussed, as well as contributing heavily to the software used to simulate the results. Thanks also to Michael Frank, Norman Margolus, Tommaso Toffoli, Tim Tyler, and Vianey Garcia-Osorio for their generous time in reviewing this work, and their helpful comments.

Portions of this work were supported by a grant from the National Science Foundation.

**Introduction**

This paper introduces a novel Reversible Cellular Automata, or RCA [note1], that we believe is computation universal, and capable of universal construction, in the sense described by Von Neumann [VonNeumann]. Where Von Neumann's original CA was irreversible, two-dimensional, and required 29 distinct states to perform its functions, the present RCA is reversible, three-dimensional, and requires only two cell states (albeit using a spatio-temporal partitioning scheme). To our knowledge, to date the simplest 2-D CA that is known to be a universal constructor is shown by Banks in [Banks]. Banks' CA was irreversible, two-dimensional, and required four distinct states. The simplest 2-D reversible cellular automata we are aware of that is known to be capable of universal computation is introduced in [Margolus], and is based on the so-called BBM (Billiard Ball Model) as discussed in [Fredkin1]. The Margolus CA is a two-state, reversible CA in two dimensions; to our knowledge it has not been ascertained whether this CA is also a universal constructor.

In this paper we will describe the general class of CA's we refer to as "Salt" automata. We then proceed to analyze one specific rule set within this new class. We will show that the CA in question is a reversible automata, and is capable of universal computation, through the existence of 'gliders' that interact to perform logic operations. We will offer strong evidence that this particular CA is also capable of so-called universal construction, through the interaction of gliders strategically emitted from a central location.

**Background and Motivation**

Research in this area has motivations beyond simple intellecutal curiosity. Moore's law cannot reasonably continue indefinitely without a move into the realm of nanoscale computational devices, and heavy reliance on three-dimensional architectures. It is becoming clear to many researchers that this transition in the physical regime within which the hardware operates will require a parallel development in software architecture and design. Specifically, there is a strong argument that it will not be possible to create densely packed nanoscale logic devices using conventional, non-reversible Boolean logic. Work by us and others in reversible computing have shown the theoretical possibility of computation using reversible logic. However, many questions remain about what overall architecture and design philosophy would best enable practical engineering work in this area to proceed.

Our approach to this design challenge is to propose an architecture that would map onto a physical implementation based on a simple substrate such as a regular crystal lattice, similar perhaps to ordinary table salt. One can imagine a system where information would be encoded through state changes at the atomic level, and would propagate through the system in response to a global synchronizing mechanism, such as pulses of electromagnetic radiation. Such an architecture has the feature that nothing in its structure is pre-ordained; a simple physical substrate is 'taught' to compute through the application of simple rules that cause local interactions at the atomic level.

We are not attempting at this point to make any claims regarding a physical realization of such a system. We are content to say that nothing we are proposing is in obvious violation of the laws of physics. With this in mind, we have attempted to develop a software-level architecture that would make such a hardware system useful. In particular, the problem of initialization of on the order of $10^{20}$ or more bits in such a device is daunting to say the least. It appears that the only feasible way to handle this problem is to have the system be capable of propagating information structures throughout the matrix, preferably at an exponential rate. This leads to a model strikingly similar to Von Neumann's original concept of self-replicating Cellular Automata [ref]. However, clearly such a physical system will not easily accomodate a complex CA with many ad-hoc rules introduced in order to effect the desired behavior. With these factors in mind, we have attempted to discover systems that are three-dimensional, reversible, require only local interaction, *and* have the attribute that the set of rules is both simple and constent enough to imagine the possibility of straightforward implementations at the molecular level.

**Previous Work**

Most of the previous work on 3D CA's has involved straightforward extensions of well-known, non-reversible CA's such as the Game Of Life. Regarding reversibility, Imai, Hori, and Morita in [Morita] show a 3D, reversible CA that is capable of a sort of reproduction, in the sense described by Langton [Langton] [Note2]. Margolus introduces a reversible, computation universal 3D lattice-gas automata in [Margolus2]. To date, we are not aware of any reversible CA that is both computation universal *and* capable of universal construction in the stronger, Von Neumann sense; nor are we aware of any three-dimensional CA, reversible or not, that has been shown to be a universal constructor. It is our intention in this paper to offer evidence that our recently discovered RCA is capable of universal construction in this stronger sense, where both the constructor and construction are themselves computation universal.

**'Salt' Automata**

The RCA in question is a member of a class of RCA's known as 'Salt' RCA's, as introduced in [Fredkin2]. Salt RCA's are three-dimensional, lattice-based cellular automata with the property that each cell is a member of one of two distinct sets, which we will call 'even' and 'odd'. For the purposes of this paper, we will refer to these two subsets collectively as 'fields'; or individually, as the 'even field' or 'odd field'. We will also refer to cells as 'even' or 'odd', indicating their membership in the corresponding field. Finally, we refer to cells as 'up' or 'down', indicating the two possible states of a cell.

A cell is even if the sum of the three integer coordinates (x, y and z) that specify its location is an even number; a cell is odd if this value is odd. The even and odd cells take up positions on the lattice that correspond to the positions of Sodium and Chloride ions in a crystal of sodium chloride, or table salt; suggesting the name, "Salt". One can visualize the Salt lattice as a three-dimensional checkerboard. Each cell has exactly twelve nearest neighbors in its own field, corresponding to the edges of a cube. Each cell also has six nearest neighbors in the opposite field, corresponding the faces of a cube. There are an additional 24 next-nearest neighbors in the opposite field, corresponding to a "knight's move" in each of the three planes perpendicular to the axes.

Evolution of a Salt automata proceeds as follows. First, we calculate a new set of changes to the values in the even field, based on the state of the odd field. Then, we update the odd field based on the state of the even field. We repeat in this way for the duration of the evolution of the automata. For the specific class of Salt models we will discuss here, the only updates allowed within a field are to conditionally swap the states of two nearest-neighbor cells (diagonal neighbors) in the same field. The decision (to swap or not to swap) is determined by the state of neighborhood cells in the opposite field. The diagonal neighbor cells that might be swapped are positioned on one of three planes, ie the XY, YZ, or XZ plane. Each cell has four diagonal neighbors in each plane, for a total of twelve neighbors, as previously described.

**The Rule**

The specific automata that we will focus on has an additional phase component, corresponding to activity in each of the three planes: a specific rule is applied in one of the three planes at each time step. The sequence of steps therefore comprises a six-phase clock, with the following operation performed at each phase:

Phase 0: apply rule to even field in XY plane
Phase 1: apply rule to odd field in YZ plane
Phase 2: apply rule to even field in XZ plane
Phase 3: apply rule to odd field in XY plane
Phase 4: apply rule to even field in YZ plane
Phase 5: apply rule to odd field in XZ plane

*Two-state, Reversible, Universal CA in 3D -- Miller, Fredkin*

The sequence chosen has the property that it is symmetric with regard to both fields and planes. In other words, any operation that is possible on a given field with an orientation with respect to a given plane, can be performed on any field and in any plane, provided you start at the appropriate step in the phase cycle.

Various rules within this set of constraints have been studied. Two in particular appear to have properties that are very intriguing from the point of view of computation universality and construction capabilities. By construction capabilities, we mean the ability for a configuration of cell states to 'construct' other configurations at points distant in the lattice, with some degree of arbitrary control over the configurations being so constructed. Ideally, such a configuration should be able to act as a machine, and be able to 'replicate' itself, by constructing another machine in an area previously empty, or filled with random or low-entropy data such as a repeating pattern (sometimes referred to as 'wallpaper').

The specific rule we will be discussing in this paper is as follows. For each diagonally situated pair of cells in a given field and given plane, we will swap their states if the following two conditions are met:

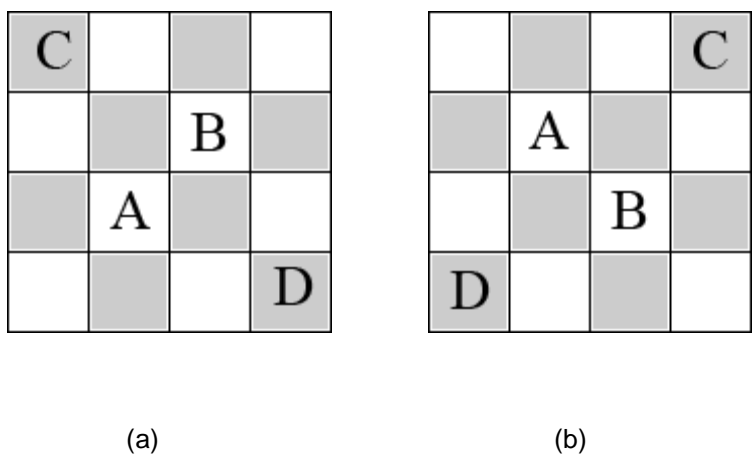

           (a)                                   (b)

*Fig. 1: Cells in positions A and B swap state iff there is an 'up' cell in positions C and/or D, and no conflicting swap possibilities. The rule is applied in both diagonal orientations, as shown in (a)*

*and (b). Odd and even fields are depicted as white and gray.*

1) there is an 'up' cell in the opposite field located at one of the two "knights move" cell locations relative to the diagonal pair, as depicted in Fig. 1.

2) There is no conflicting swap possible for either cell in the pair.

This simple rule turns out to have some interesting properties. First, it is easy to construct a "glider", or configuration that can translate itself in one of several possible directions. The simplest glider consists of two cells of opposite parity, situated in the knight's move orientation, as depicted in Fig. 2.

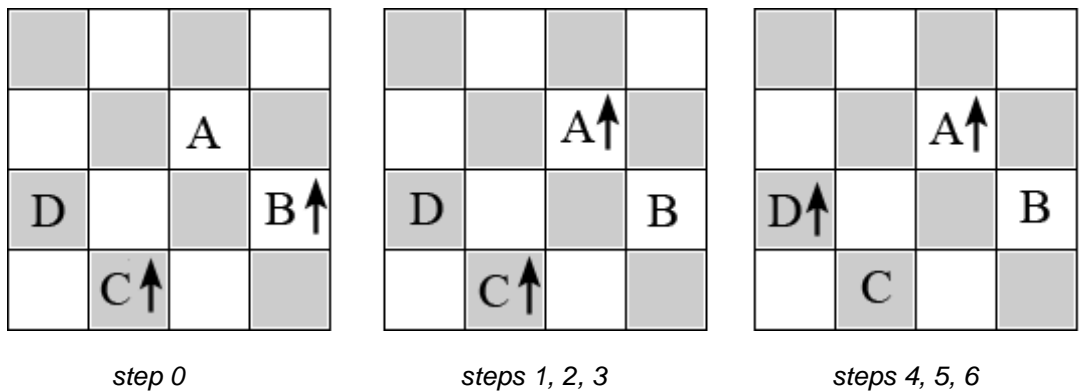

*step 0*  *steps 1, 2, 3*  *steps 4, 5, 6*

*Fig. 2: The simplest glider configuration. 'up' cells have up-pointing arrows; 'down' cells are left blank. At the transition from step 0 to step 1, cells A and B swap states due to cell C being 'up'. At the transition from step 3 to step 4, cells C and D swap states due to cell A being 'up'. At this point the glider has positioned itself at a new location, one cell up and one to the left of the original configuration. In future steps, this glider will continue to move up and to the left.*

Gliders of this type can be constructed to move in any of the twelve directions corresponding to diagonal movement in one of four directions, in each of the three planes.  It is worth noting that there are two possible two-cell gliders for each of the twelve directions: one with the odd cell on the left in the direction of motion, and one with the even cell on the left.  This gives us a total of 24 possible two-cell glider configurations.  Their available directions of motion can be visualized as vectors from the center of a cube, pointing towards the midpoint of each of the cube's twelve edges.

**Reversibility**

To show that this CA is in fact an RCA (ie, Reversible Cellular Automaton), we need to show that for every state n and subsequent state n+1, there is a unique, one-to-one onto mapping.  Another way to describe this requirement is to say that for all n, state n+1 is uniquely determined by state n, and state n is uniquely determined by state n+1.

Noting that the rule simply swaps the states of certain pairs of cells in the even field, depending on the state of affairs in the odd field, we can make the following obvious assertion:  If we repeat the same phase-conditional swapping operation on the even field, based on the state of the odd field, the even field will be restored to its original state, the state that existed before the first set of swaps took place.  This is a consequence of the fact that a swapping operation is its own inverse.

If we now continue in reverse, swapping the odd field depending on the state of the even field (using the phase rule that was used originally; thus the order in which we operate on each plane and field is reversed), we return the odd field to the state it was in before its last swap. Continuing this set of operations allows us to run the evolution of the CA backwards; each previous step is completely determined by the present state of the two fields.  Therefore, the Salt CA is in fact an RCA.

**Signals And Routing**

The paradigm we will employ to describe how this RCA can achieve computation is similar in spirit to the "Billiard Ball Model" (BBM), as described by Fredkin and Toffoli in [Fredkin1]. In the discussion that follows, signals will be represented by gliders. At specific time steps, the existence of a glider at a specific location will indicate a logic "1", or True. Absence of a glider at this location and time indicates a logic "0", or False.

The first requirement we have is to show that signals can be routed arbitrarily from point to point. Our simple two-cell glider's direction of travel can be changed by the strategic placement of a single cell, which is not affected by the glider. We will refer to such cells as reflectors. Fig. 3 shows a glider being reflected. Note that the reflection causes the glider to take up a new orientation in a different plane.

Combinations of reflectors can redirect a glider onto a path in any of the twelve possible directions.

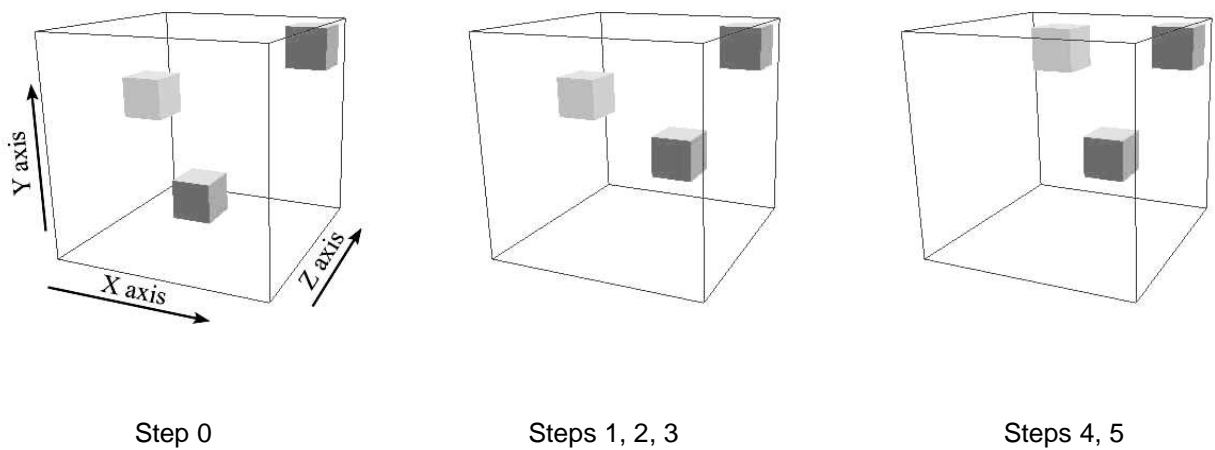

Step 0             Steps 1, 2, 3             Steps 4, 5

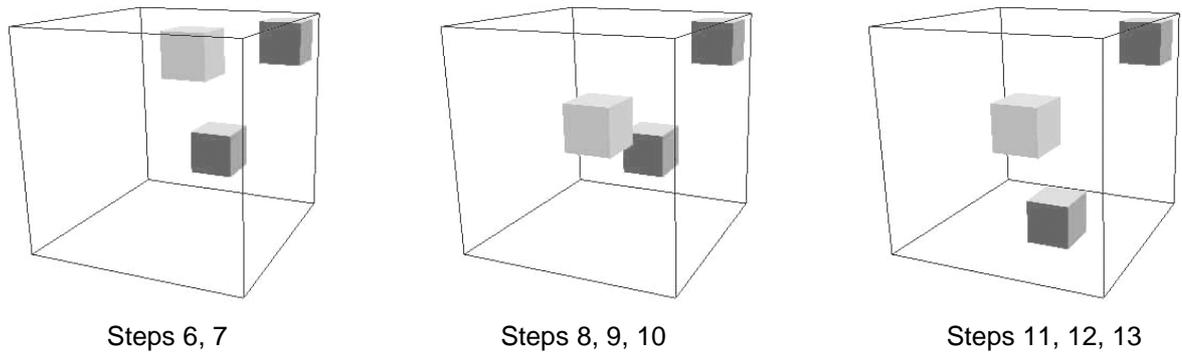

| Steps 6, 7 | Steps 8, 9, 10 | Steps 11, 12, 13 |

*Fig. 3: A glider moving up and to the right is deflected into another plane. Up, even cells are light gray; up, odd cells are dark gray. Down cells are not shown. Initially, the glider is traveling in the +X, +Y direction. After deflection, the glider is traveling in the -Y, -Z direction.*

**Logical Operations and Universal Computation**

Again following the BBM model, we will show that the Salt RCA is capable of being configured so as to perform arbitrary logic operations on inputs, with controllable delays. It has been shown that any CA that can meet this requirement can, given infinite memory, perform Turing-complete computation [Fredkin1]. In the case of finite memory limits, a CA with this capability can emulate the operation of any other finite, arbitrary arrangement of logic gates, using a finite number of cells. In particular, such a CA can for instance emulate the behavior of a Pentium-based computer, or any other general-purpose computational device.

We will show that the Salt RCA can duplicate the function of a series of interconnected "Interaction Gates", as discussed in [Fredkin1] and [Margolus]. An interaction gate can be depicted diagrammatically as in fig. 4.

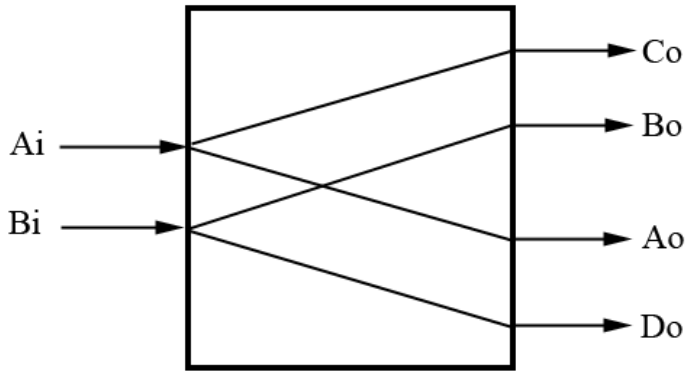

*Fig. 4: The Interaction Gate. Inputs Ai and Bi arrive at respective outputs Ao and Bo, <u>unless</u> both inputs arrive at the same time (ie, both are logic "1"), in which case two 1's are output at Co and Do.*

The truth table for the Interaction Gate is as follows:

```
Ai Bi | Ao Bo Co Do
-------------------
0  0  | 0  0  0  0
0  1  | 0  1  0  0
1  0  | 1  0  0  0
1  1  | 0  0  1  1
```

It is apparent that both outputs Co and Do comprise the function of logical AND. Logical NOT can be obtained by supplying input Bi with a logic "1" at every time step; in this case, the output Bo is equivalent to NOT Ao.

FANOUT (duplication of a signal to multiple outputs) is achieved by again providing input Bi with a stream of 1's, producing a duplicate of input Ai at outputs Co and Do.  Clearly, arbitrary boolean logic can be achieved simply by connecting the inputs and outputs of Interaction Gates, and supplying some inputs with a stream of logical 1's.  One may object that in practice such an approach will generate increasing amounts of "garbage", in the form of outputs that are not needed for further computation.  Much work has been done to show that this is not in fact the case; in general, arbitrary computational problems can be solved with reversible, conservative logic of this type, with at most a fixed proportional increase in the number of gates and wires that would be required by traditional, irreversible logic [Fredkin1].  This increase is on the order of 1.5 to 2 times the gate count of a conventional circuit.

Our implementation of the Interaction Gate is similar in spirit to that used for the BBM model.  Simply put, two pathways are defined where gliders representing logical 1's may travel.  The pathways intersect in such a way that if both gliders are traveling down their respective paths at the same time, they will interact so as to deflect each other's direction onto two alternate pathways.  There are four possibilities, corresponding to the entries in the truth table for the Interaction Gate.

In the first case, both inputs are logic "0"; no gliders are present, so of course all four outputs are also zero.

In the second and third case, a logic "1" (represented by a glider) travels down either the pathway from Ai to Ao, or Bi to Bo.  In these cases, either Ao or Bo produce a logic "1", and all other outputs are zero.

In the fourth and final case, shown pictorially in Fig 5, both inputs are logic "1", meaning both pathways have gliders.  In this case, the interaction between the two gliders causes their paths to be deflected to the output paths corresponding to outputs Co and Do in the truth table.

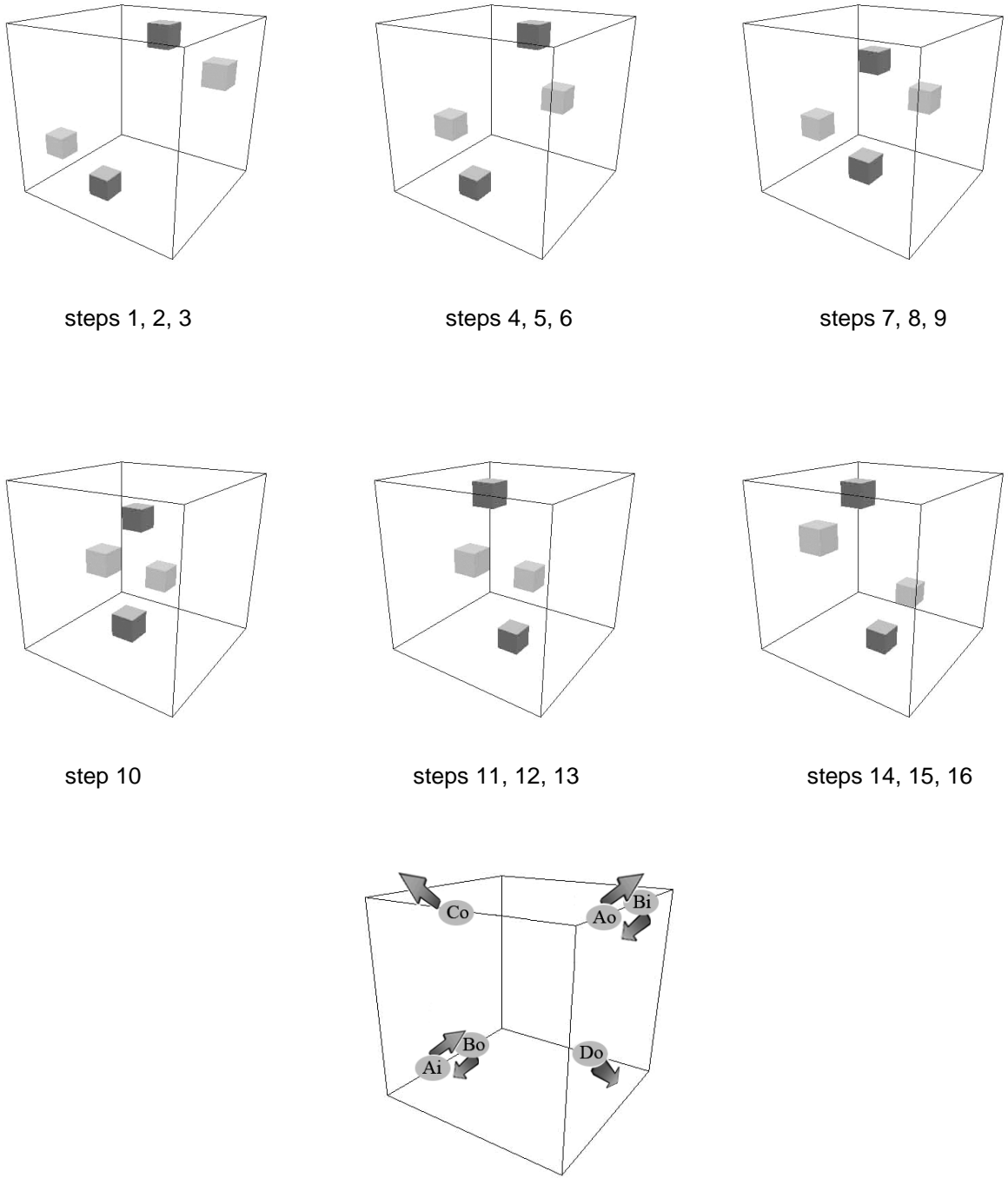

*Fig. 5: Two gliders form an Interaction Gate. If both gliders are present, as shown here, they deflect each other onto two different paths than either would have followed alone. The gray labels on the bottom image show a cubic region of cells representing an Interaction Gate, with the approximate locations of inputs and outputs corresponding to the functional diagram shown in Figure 4.*

**Universal construction**

We have shown that our RCA is a RUCA; it can duplicate the functioning of any interconnection of logic gates, and therefore, resources permitting, any Turing-complete computational system of any kind. Now we turn to the problem of construction. We define "construction" as the ability for a configuration of cells to establish a second specific configuration of cells within the lattice, at a point distant from the configuration itself. There are various subtleties discussed in the literature, in formally establishing the criterion by which a CA can be judged capable of so-called universal construction. For the purposes of this paper, we will assume the following definitions, which we believe are equivalent to Von Neumann's:

An **Arbitrary Machine** is a configuration of cells within a Cellular Automata that is capable of reproducing the behavior of an arbitrarily connected series of Boolean Logic Gates and delays. Such reproduction may occur at any specific timescale; ie, 100 steps of the CA may equal one 'timestep' of the Boolean network. We assume such an Arbitrary Machine has a non-exhaustible supply of gliders as at least one input, and is allowed to produce outputs that are not relevant to any particluar task (so-called 'garbage' outputs).

A Cellular Automata is capable of **universal construction** iff the following conditions are met:

1) The CA in question is capable of supporting Arbitrary Machines.

2) An Arbitrary Machine can be established in the CA in question that is capable of itself constructing an identifiably distinct Arbitrary Machine at a location in the lattice that was previously filled with random data, or low-entropy data such as a repetitive pattern.

3) There are in general no limits to the size or complexity of the Arbitrary Machines so constructed.

While we are not ready to prove universal construction is possible with the Salt RUCA, we believe there is ample evidence to suggest that this is in fact the case. We will outline our reasons for believing this is so.

In the discussion that follows, we are assuming that construction proceeds in a lattice that is primarily filled with 'down' cells. Our initial Arbitrary Machine is a configuration of 'up' cells, with the additional resource of a non-exhaustible supply of gliders to be used both as logic 1's and as building materials for the construction of other machines.

The first requirement we have is that we be able to position a single cell at an arbitrary point in the lattice, from a remote location. While there are several ways to do this, one method which appears most promising involves the introduction of a new kind of glider, a 4-cell configuration that proceeds at one third the speed of the simple, 2-cell glider previously introduced. The operation of this glider is depicted in Fig. 6. This type of 4-cell glider can be created through the interaction of three 2-cell gliders [note3].

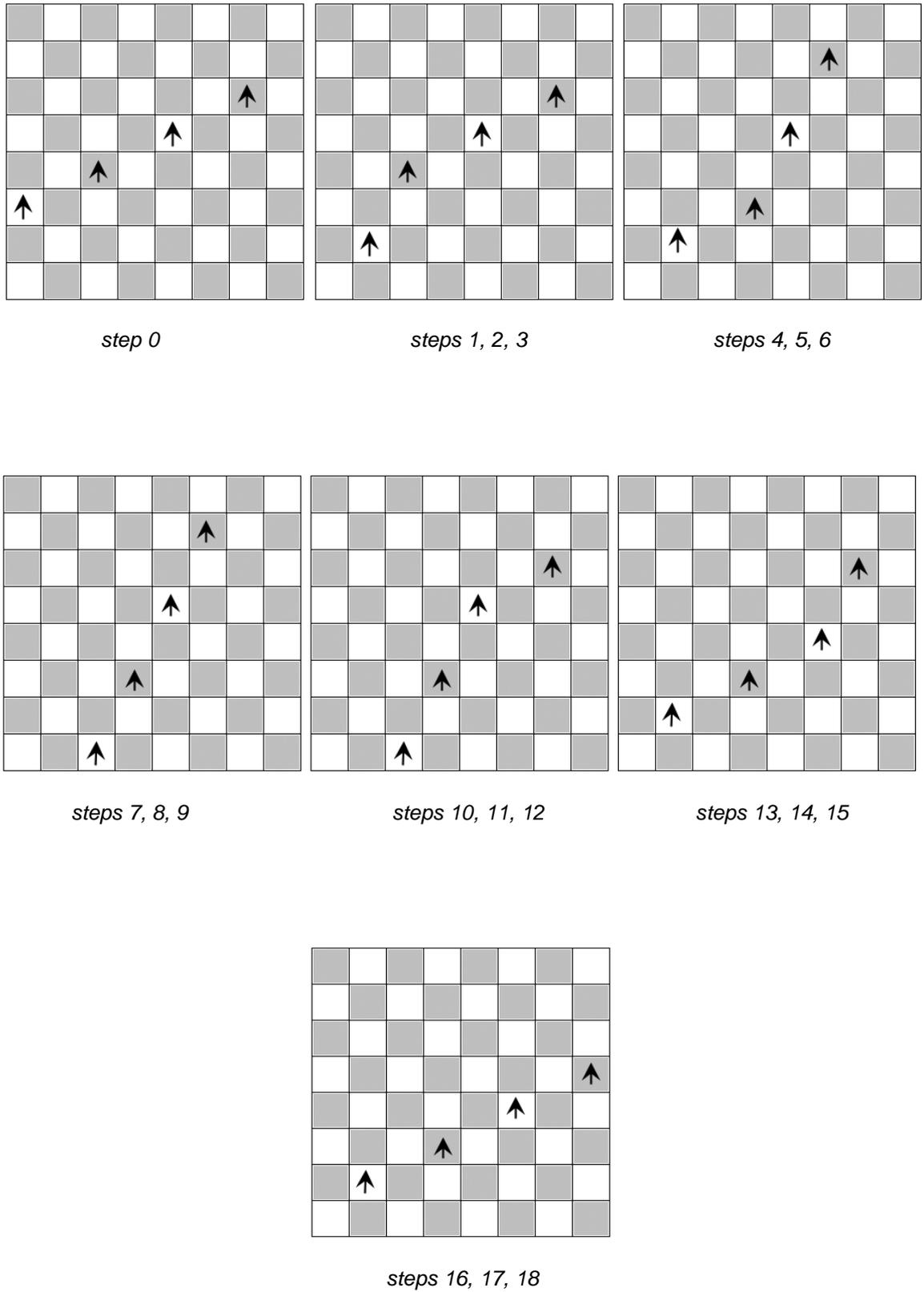

Fig 6: A 4-cell glider, depicted in the XY plane. This glider moves one cell diagonally every 18

*steps, or 1/3 the speed of the 2-cell glider. Note how the 'no conflicting swaps' rule affects the behavior of some of the cells. While the outermost cells swap on every state, the innermost two cells are unable to swap in 2 out of 3 possible cases, due to conflicting swap possibilities indicated by their two nearest neighbors.*

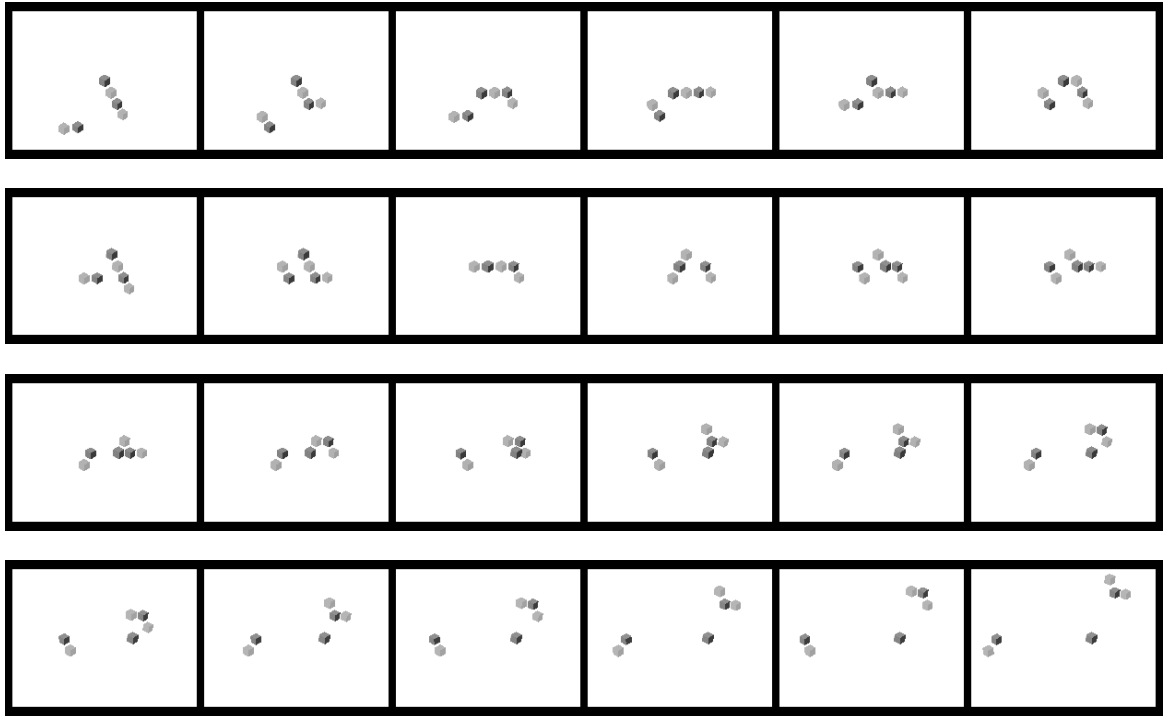

*Fig 7: A fast and a slow glider moving in the same direction. The fast glider overtakes the slow one; after the interaction, a single cell remains. One of the gliders is deflected off the original path; the other loses a cell, but continues along the original trajectory. (Only steps where the configuration changes are shown; the camera angle is along a 3-axis diagonal. Sequence proceeds from left to right.)*

*Operations such as this can be used both to position cells in precise locations, and to produce gliders moving along trajectories not available from a centrally located emitter.*

In Fig. 7, we show one possible interaction between a slow glider and a fast glider. In this case, the two gliders are travelling in the same direction (the fast glider is positioned one cell deeper in the Z plane). At the point where the fast glider catches up with the slower one, we see an interaction that eventually results in a fast 2-cell glider moving at an altered angle; a 3-cell glider moving in the original direction of the 4-cell glider (3 cell gliders move at the same speed as 2-cell gliders); and a single, nonmoving cell near the point of interaction.

This operation has a dual use: it can position a cell in a precise location from a distant 'emitter' configuration; and it can also be used to effect a 2-cell glider travelling along a trajectory that would not be possible coming directly from the emitter. In effect, interactions such as this can be used to progam gliders to perform operations such as "go 100 cells in the X,Y direction, and then emit a glider in the Y,Z direction."

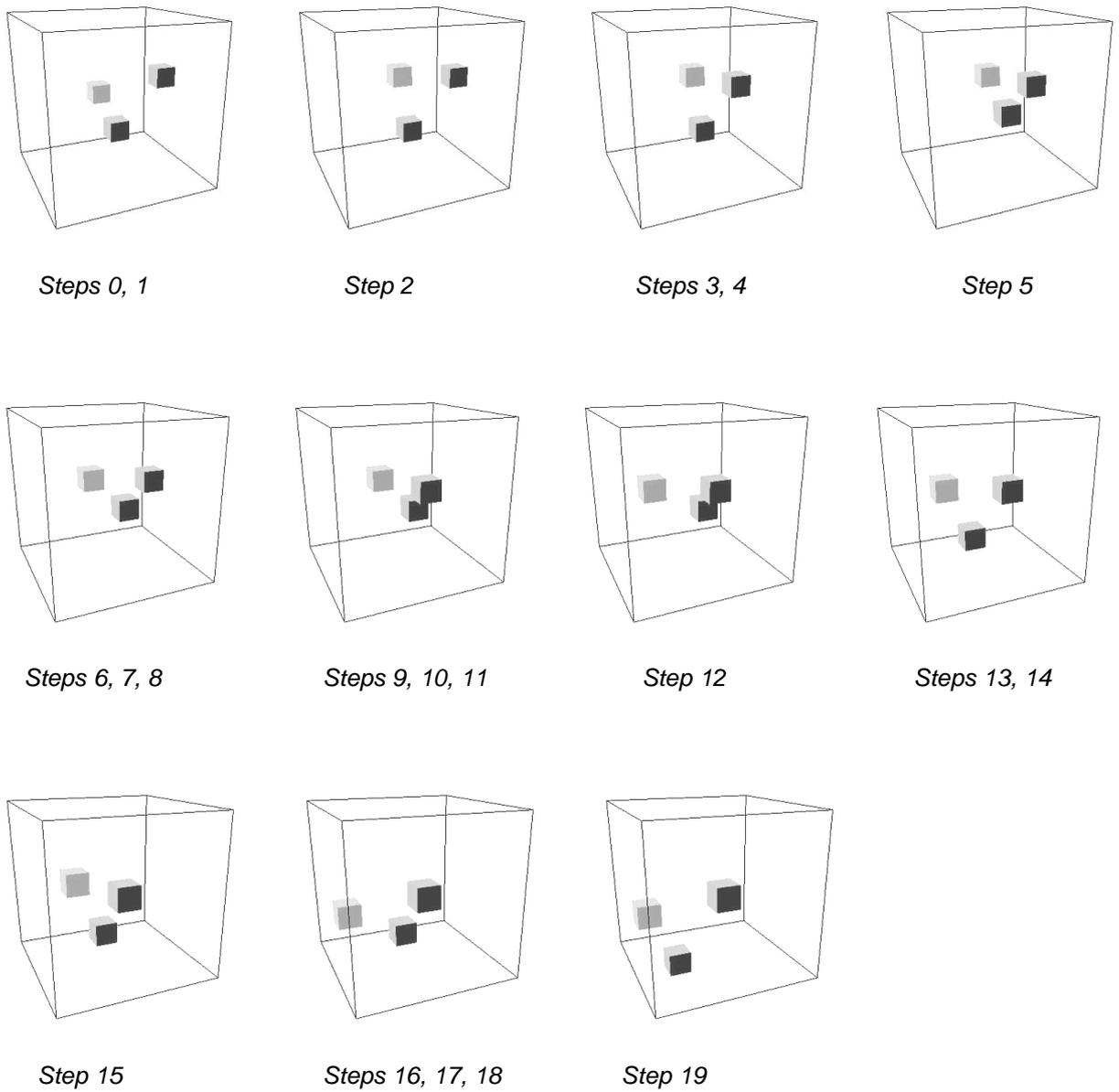

*Fig 8: A 2-cell glider repositions a single cell*

Single cells can also be repositioned through a simple interaction with a 2-cell glider, as depicted in Fig. 8.

Complex combinations of such directives should enable a sophisticated Arbitrary Machine to be able to orchestrate a series of gliders converging on a central location. Such interactions should be sufficient to allow the creation of Arbitrary Machines at distant locations.

**Physical Implementation**

Recent research has indicated that the Salt RUCA might be realizable as a physical device at the nanometer scale. Specifically, we are investigating the possibility that a physical crystal structure, similar in some respects to ordinary table salt ($Na^+Cl^-$), might be induced to follow the Salt RUCA rules, where the energy or spin states of individual atoms would represent the states of cells in the automaton. This work is being conducted under a grant by the National Science Foundation.

**Conclusion**

We have introduced a novel, two-state, three-dimensional RUCA that is clearly capable of universal computation. We have introduced evidence that this RUCA is also a universal constructor, in the sense described in [VonNeumann]. Future work would involve running further simulations, working out the details of computation and construction within this class of Salt RUCA's. It should be possible to show conclusively that universal construction is possible within the Salt RUCA framework.

*Notes:*

[note1] *For a comprehensive review of cellular automata theory, see [Ilachinski]. In particular, pages 369 - 385 include an excellent introduction to the history and significance of reversible CA's.*

[note2] *The 'XOR' rule, first mentioned in Martin Gardner's* Scientific American *column in 1971 (reprinted in [Gardner]), is an early example of a trivially simple CA that is nontheless capable of reproduction in this somewhat limited sense.*

[note3] *Reflection on the nature of reversible CA's will make clear that no interaction between two 2-cell gliders can form a 4-cell glider. If this were possible, the scenario would have to be reversible. Reversing the action of a 4-cell glider causes it to move in the opposite direction indefinitely. Since such a glider will never spontaneously break down into two 2-cell gliders, no two 2-cell gliders could have combined to produce the 4-cell glider. In the three-glider interaction, the disposition of the remaining 2-cell glider 'encodes' information necessary to recreate the interaction in reverse. Therefore the reverse interaction is also possible: a 2-cell glider can break a 4-cell glider into two 2-cell gliders.*

*Two-state, Reversible, Universal CA in 3D -- Miller, Fredkin*